\documentstyle{article}
\font\eurm=eurm10       
\newcommand{\Eu}[1]{\mbox{\eurm #1}}

\font\msam=msam10 scaled 900

\font\msbm=msbm10 scaled 900
\font\msbm=msbm10
\font\msbmsmall=msbm10 scaled 800
\font\msbmtiny=msbm10 scaled 700
\newfam\msbmfam
\textfont\msbmfam=\msbm
\scriptfont\msbmfam=\msbmsmall
\scriptscriptfont\msbmfam=\msbmtiny
\def\Bbb{\fam\msbmfam\msbm}

\def\Re{\mbox{\rm Re}\,}

\def\Im{\mbox{\rm Im}\,}
\def\varkappa{\mbox{\msbm\char'173}}      

\newtheorem{prop}{Proposition}[section]

\newtheorem{theorem}[prop]{Theorem}

\def\proof{\par{\it Proof}. \ignorespaces}
\def\endproof{{\msam \char'3}\par\smallskip}

\begin{document}

\centerline{\Large Quasi-linear Stokes Phenomenon for}
\centerline{\Large the Second Painlev\'e Transcendent}

\bigskip
\centerline{\large A.~R.~Its,}
\centerline{\it Department of Mathematical Sciences,}
\centerline{\it Indiana University -- Purdue University  Indianapolis}
\centerline{\it Indianapolis, IN 46202-3216, USA}

\centerline{\large A.~A.~Kapaev,}
\centerline{\it St Petersburg Department of Steklov Mathematical Institute,}
\centerline{\it Fontanka 27, St Petersburg 191011, Russia}

\begin{abstract}
Using the Riemann-Hilbert approach, we study the quasi-linear Stokes
phenomenon for the second Painlev\'e equation $y_{xx}=2y^3+xy-\alpha$. The
precise description of the exponentially small jump in the dominant solution
approaching $\alpha/x$ as $|x|\to\infty$ is given. For the asymptotic power
expansion of the dominant solution, the coefficient asymptotics is found.
\end{abstract}

\section{Introduction}

The classical Painlev\'e equations \cite{ince} are six particular ODEs
$y_{xx}=R(x,y,y_x)$ whose general solutions called the Painlev\'e functions
are free from movable critical points and can not be expressed in terms of
elementary or classical special functions. The properties of the Painlev\'e
functions show the remarkable similarity to those of the classical
transcendents of hypergeometric type like Airy, Bessel, Weber, Whittaker and
Gauss functions, see monograph \cite{its_nov} and surveys mentioned there.

The problem of the asymptotic description of the Painlev\'e functions near
fixed critical points is one of the most interesting and important problems
in the theory of the Painlev\'e equations. The first result of such kind was
obtained by Boutroux \cite{boutroux}, who studied the first and second
Painlev\'e equations, $y_{xx}=6y^2+x$ and $y_{xx}=2y^3+xy-\alpha$, using
direct methods based on the integral estimates. Boutroux has shown that,
generically, the asymptotic as $|x|\to\infty$ behavior of the first and
second Painlev\'e functions is given by the Weierstra\ss\ elliptic function
and by the elliptic sine, respectively. Periods of these elliptic asymptotic
solutions depend on $\arg x$ only. Along the rays
$\arg x=\pi+\frac{2\pi}{5}k$ for P1 and $\arg x=\frac{\pi}{3}k$ for P2,
$k\in{\Bbb Z}$, the general Painlev\'e transcendent is described by the
degenerate elliptic, i.e.\ trigonometric, functions. The asymptotic forms of
the phase shift in the elliptic ansatz for P1 and P2 as $\arg x$ belongs to
the interior of the ``elliptic" sectors are found in
\cite{joshi_kruskal, nov}, while the connection formulae for the phase shift
for different sectors were obtained in \cite{kapaev:ell, kap_kit}. Similar
results for Painlev\'e equations P4 and P3 were published in
\cite{kapaev4, nov2}. The nontrivial jump in the phase of the elliptic
asymptotic ansatz across the rays of the trigonometric asymptotic behavior
constitutes the {\it nonlinear} Stokes phenomenon.

In our terms, the {\it quasi-linear} Stokes phenomenon is the analog of the
classical, or linear, Stokes phenomenon. The latter was observed for the
first time in the asymptotic analysis of the Airy equation, $y_{xx}=xy$, and
consists in the jump of the recessive, exponentially decreasing solution in
the ``shadow" of the dominant, exponentially large term.

Boutroux \cite{boutroux} has found that the Painlev\'e equations admit
1-parameter solutions represented by the sum
\begin{equation}\label{gen_struct}
y=\hbox{(power series)}+\hbox{(exponential terms)}
\end{equation}
of a power series with the leading term $y_0(x)$ which satisfies certain
algebraic equation, e.g.\ $6y_0^2+x=0$ for P1 and $2y_0^3+xy_0=\alpha$ for
P2, and of a {\it transseries} which is the sum of decreasing exponential
terms. The latter can be obtained from the conventional perturbation
analysis. Solutions of such kind are usually referred to as the truncated,
separatrix, degenerate, quasi-stationary or instanton solutions.

Since all equations of the perturbation analysis are linear homogeneous and
non-homogeneous ODEs, there is no surprise to meet the quasi-linear Stokes
phenomenon studying the degenerate solutions.

The first, heuristic, description of the quasi-linear Stokes phenomenon was
obtained in \cite{kapaev:P1, kapaev:P2}. There are known several attempts to
find the analytic meaning of the quasi-linear Stokes phenomenon and to
justify it using the resurgent analysis. The reader can find general settings
of the Borel summation technique for the nonlinear differential equations in
the paper of Costin \cite{costin1}. The technique is shown to be effective
for certain nonlinear asymptotic problems. For instance, in \cite{costin2},
Costin applied the resurgent approach to find the asymptotic location of the
first real pole of the degenerate solution of P1 when the amplitude of its
exponentially decreasing perturbation becomes large.

Takei \cite{takei} proposed two ways of the use of the Borel summation for
description of the quasi-linear Stokes phenomenon in P1. His first idea is
similar to one of \cite{costin1} and is based on the study of the Borel
transform of the leading power series. Assuming that the Borel series has
certain analytic structure on the boundary of its convergence, Takei obtained
a simple formula which relates the jump of the recessive term in
(\ref{gen_struct}) to the asymptotics of the Taylor coefficients of the Borel
transform. However, the assumptions on the analytic structure of the Borel
transform remain unjustified. In the second way, he studied the linear
differential equation associated to P1 \cite{JMU} using the so-called
``exact" WKB analysis. This potentially effective approach is based on the
assumption of the Borel summability of the formal ``instanton" series for the
Painlev\'e function which is still an open problem, see \cite{takei2}.

The coefficient asymptotics in the power series in (\ref{gen_struct}) for the
fifth Painlev\'e equation P5 was found by Basor and Tracy \cite{BT} up to a
common constant factor. The approximate value for the latter was proposed in
\cite{BT} using the numerical computations. The exact value of the
Basor-Tracy constant as well as the proof of their asymptotic formula were
presented by Andreev and Kitaev in \cite{AK} via a combination of the
isomonodromy method and the Borel summation technique. We emphasize the
importance of the paper \cite{AK} as the first exact description of the
relation between the coefficient asymptotics in the Painlev\'e transcendent
expansions and the quasi-linear Stokes phenomenon.

In \cite{JK}, Joshi and Kitaev studied the asymptotics of the leading power
series coefficients for the degenerate solution of P1 and announced some
results in the asymptotics of the transseries coefficients. The investigation
is based on the use of the recursion relations and leaves a common factor for
all the asymptotic coefficients undetermined. The authors proposed the value
of this factor noting that it could be found using the quasi-linear Stokes
phenomenon and the Borel transform technique similar to presented in
\cite{AK}.

The common idea of the approaches mentioned above is the use of the Borel
transform, most frequently, to fix the solution having the formal power
series as its asymptotics. The problem with this idea is that one needs to
assume certain analytic properties of the Borel transform of the solution.

Below, we study the quasi-linear Stokes phenomenon for the second Painlev\'e
equation P2,
\begin{equation}\label{P2}
y_{xx}=2y^3+xy-\alpha,\quad
\alpha=const,
\end{equation}
from the viewpoint of the isomonodromy deformation method
\cite{JMU, FN, its_nov}. We pursue a two-fold goal, i.e.\ (a)~to make an
exact sense of the formal equation (\ref{gen_struct}), and (b)~to evaluate
the asymptotics of the coefficients of the leading power series in
(\ref{gen_struct}).

We stress that we {\it do not\/} use the Borel summation at any stage of our
investigation. Instead, we fix the solutions of P2 (\ref{P2}) with the power
series asymptotic behavior by the proper choice of the monodromy data of the
associated linear equation. Unlike \cite{AK}, our study is based on the
direct asymptotic analysis of the Riemann-Hilbert problem via the nonlinear
steepest descent method \cite{DZ}. In this approach, we associate the
solution to the RH problem with a disjoint graph. This in turn allows us to
separate effectively the exponentially small term from the power-like
background, see Section~2. Using the outlined method, we rigorously prove the
existence of the solutions $y_1(x)$, $y_2(x)$, $y_3(x)$ asymptotic to
$\alpha/x$ as $|x|\to\infty$ in the respective overlapping sectors
$\arg x\in(-\frac{\pi}{3},\pi)$, $\arg x\in(\frac{\pi}{3},\frac{5\pi}{3})$,
$\arg x\in(-\pi,\frac{\pi}{3})$, and find the exponentially small differences
$y_k-y_l$, $k,l=1,2,3$, which constitute the quasi-linear Stokes phenomenon.

The collection of the functions $y_k(x)$, $k=1,2,3$, forms a piece-wise
holomorphic function $\hat y(x)\sim\frac{\alpha}{x}$ as $|x|\to\infty$.
The moments of this function immediately yield the asymptotics for the
coefficients of the leading power series (\ref{gen_struct}).

\section{Riemann-Hilbert problem for P2}

In \cite{FN}, Flaschka and Newell formulated the inverse problem method for
P2. The method reduces the integration of (\ref{P2}) for
$|\Re\alpha|<\frac{1}{2}$ to solution of the matrix Riemann-Hilbert (RH)
problem for an auxiliary function $\Psi(\lambda)$ which in its turn is
equivalent to a system of singular integral equations. Using this method,
they described the limit of small $y$ and $y_x$ and obtained the determinant
formulae for the rational and classical solutions of (\ref{P2}). Later, the
RH problem was extensively studied by Fokas, Ablowitz, Deift, Zhou, Its,
Kapaev, who considered the questions of existence and uniqueness of the RH
problem solution \cite{FA, FZ}, and obtained various asymptotic solutions of
the latter \cite{DZ, IFK}.

In accord with \cite{FN}, the Painlev\'e function set is parameterized by
two of the Stokes multipliers of the associated linear system below denoted
by the symbols $s_k$, $k=1,2,3$. We begin with the case of decreasing as
$x\to+\infty$ Painlev\'e transcendents corresponding to the special value
$s_2=0$. We observe that, for $0<\arg x<\frac{\pi}{3}$, the RH problem graph
can be transformed to the union of three disjoint branches. Using this fact,
we define the dominant term $y_1(x)$ as the solution of the Painlev\'e
equation corresponding to the reduced RH problem with $s_2=s_3=0$. The
recessive, exponentially small term appears as the contribution of the
disjoint branches in the RH problem graph. Thus the recessive term can be
effectively distinguished from the dominant background.

Below, we use the modification of the original RH problem of Flaschka and
Newell \cite{FN} which is more convenient for our purposes. Also, we restrict
ourselves to the case $\alpha-\frac{1}{2}\notin{\Bbb Z}$. The excluded case
of the half-integer $\alpha$ can be treated similarly, but requires a separate
consideration due to logarithmic behavior of $\Psi(\lambda)$ at the origin.

Let us introduce the piece-wise oriented contour
$\gamma=C\cup\rho_+\cup\rho_-\cup_{k=1}^6\gamma_k$ which is the union of the
rays $\gamma_k=\{\lambda\in{\Bbb C}\colon\
|\lambda|>r,\
\arg\lambda=\frac{\pi}{6}+\frac{\pi}{3}(k-1)\}$, $k=1,\dots,6$, oriented to
infinity, the clock-wise oriented circle $C=\{\lambda\in{\Bbb C}\colon\
|\lambda|=r\}$, and of two vertical radiuses
$\rho_+=\{\lambda\in{\Bbb C}\colon\ |\lambda|<r,\
\arg\lambda=\frac{\pi}{2}\}$
and
$\rho_-=\{\lambda\in{\Bbb C}\colon\ |\lambda|<r,\
\arg\lambda=-\frac{\pi}{2}\}$
oriented to the origin. The contour $\gamma$ divides the complex
$\lambda$-plane into 8 regions $\Omega_k$, $k\in\{l,r,1,\dots,6\}$.
$\Omega_l$ and $\Omega_r$ are left and right halves of the interior of the
circle $C$ deprived the radiuses $\rho_+$, $\rho_-$. The regions $\Omega_k$,
$k=1,\dots,6$, are the sectors between the rays $\gamma_k$ and $\gamma_{k-1}$
outside the circle.

Let each of the regions $\Omega_k$, $k=r,6,1,2$, be a domain for a
holomorphic $2\times2$ matrix function $\Psi_k(\lambda)$. Denote the
collection of $\Psi_k(\lambda)$ by $\Psi(\lambda)$,
\begin{equation}\label{Psi_collect}
\Psi(\lambda)\bigr|_{\lambda\in\Omega_k}=\Psi_k(\lambda),\quad
\Psi(e^{i\pi}\lambda)=\sigma_2\Psi(\lambda)\sigma_2.
\end{equation}
Let $\Psi_+(\lambda)$ and $\Psi_-(\lambda)$ be the limits of $\Psi(\lambda)$
on $\gamma$ to the left and to the right, respectively. Let us also introduce
the Pauli matrices $\sigma_3=\bigl({\hskip-2pt1\ \ 0\atop0\!\ -1}\bigr)$,
$\sigma_2=\bigl({0\ -i\atop i\ \ 0}\bigr)$,
$\sigma_1=\bigl({0\ 1\atop1\ 0}\bigr)$ and two matrices
$\sigma_+=\bigl({0\ 1\atop0\ 0}\bigr)$,
$\sigma_-=\bigl({0\ 0\atop1\ 0}\bigr)$. The RH problem we talk about is the
following one:
\begin{description}
\item{i)} Find a piece-wise holomorphic $2\times2$ matrix function
$\Psi(\lambda)$ such that
\begin{equation}\label{p28}
\Psi(\lambda)e^{\theta\sigma_3}\to I,\quad
\lambda\to\infty,\quad
\theta=i\bigl(\frac{4}{3}\lambda^3+x\lambda\bigr),
\end{equation}
and
\begin{equation}\label{Psi_at_0}
\|\Psi_r(\lambda)\lambda^{-\alpha\sigma_3}\|\leq const\quad\hbox{as}\quad
\lambda\to0;
\end{equation}
\item{ii)} on the contour $\gamma$, the jump condition holds
\begin{equation}\label{p29}
\Psi_+(\lambda)=\Psi_-(\lambda)S(\lambda),
\end{equation}
where the piece-wise constant matrix $S(\lambda)$ is given by
equations:\hfill\par
\noindent
on the rays $\gamma_k$,
\begin{equation}\label{p23}
S(\lambda)\bigr|_{\gamma_k}=S_k,\quad
S_{2k-1}=I+s_{2k-1}\sigma_-,\quad
S_{2k}=I+s_{2k}\sigma_+,
\end{equation}
with the constants $s_k$ satisfying the constraints
\begin{equation}\label{p24}
s_{k+3}=-s_k,\quad
s_1-s_2+s_3+s_1s_2s_3=-2\sin\pi\alpha;
\end{equation}
on the radiuses $\rho_{\pm}$, $S(\lambda)$ is specified by the equations
\begin{eqnarray}\label{M_rels}
\lambda\in\rho_-\colon\
&&\Psi_l(e^{2\pi i}\lambda)=\Psi_r(\lambda)M,\nonumber
\\
\lambda\in\rho_+\colon\
&&\Psi_r(\lambda)=\Psi_l(\lambda)\sigma_2M\sigma_2,
\end{eqnarray}
where
\begin{equation}\label{M}
M=ie^{i\pi\alpha\sigma_3}\sigma_1;
\end{equation}
\noindent
on the circle $C$, the piece-wise constant jump matrix $S(\lambda)$ is
defined by the equations
\begin{eqnarray}\label{E_rels}
&&\hskip-23pt
\Psi_6(\lambda)=\Psi_r(\lambda)ES_6^{-1},\quad
\Psi_1(\lambda)=\Psi_r(\lambda)E,\quad
\Psi_2(\lambda)=\Psi_r(\lambda)ES_1,
\\
&&\hskip-23pt
\Psi_3(\lambda)=
\Psi_l(\lambda)\sigma_2E\sigma_2S_3^{-1},\
\Psi_4(\lambda)=\Psi_l(\lambda)\sigma_2E\sigma_2,\
\Psi_5(\lambda)=\Psi_l(\lambda)\sigma_2E\sigma_2S_4,\nonumber
\end{eqnarray}
where the unimodular constant matrix $E$ satisfies the equation
\begin{equation}\label{E_conditions}
ES_1S_2S_3=\sigma_2M^{-1}E\sigma_2.
\end{equation}
\end{description}

Because the asymptotics of $\Psi(\lambda)$ as $\lambda\to\infty$ is given by
\begin{equation}\label{Y_expansion}
Y(\lambda):=\Psi(\lambda)e^{\theta\sigma_3}=
I+\frac{1}{\lambda}\bigl(-\frac{iD}{2}\sigma_3+\frac{y}{2}\sigma_1\bigr)+
{\cal O}(\frac{1}{\lambda^2}),\quad
D=y_x^2-y^4-xy^2+2\alpha y,
\end{equation}
the solution $y(x)$ of the Painlev\'e equation can be found from the
``residue" of $Y(\lambda)$ at infinity,
\begin{equation}\label{y_from_Y}
y=2\lim_{\lambda\to\infty}\lambda Y_{12}(\lambda)=
2\lim_{\lambda\to\infty}\lambda Y_{21}(\lambda).
\end{equation}
The equation (\ref{y_from_Y}) specifies the Painlev\'e transcendent as the
function $y=f(x,\alpha,\{s_k\})$ of the independent variable $x$, of the
parameter $\alpha$ in the equation and of the Stokes multipliers $s_k$ (in
the generic case, the connection matrix $E$ can be expressed via $s_k$ using
the equation (\ref{E_conditions}) modulo the left diagonal multiplier). Using
the solution $y=f(x,\alpha,\{s_k\})$ and the symmetries of the Stokes
multipliers described in \cite{kapaev:ell}, we obtain further solutions of P2:
\begin{eqnarray}\label{P_symmetries}
&&y=-f(x,-\alpha,\{-s_k\}),\quad
y=\overline{f(\bar x,\bar\alpha,\{\overline{s_{4-k}}\})},\nonumber
\\
&&y=e^{i\frac{2\pi}{3}n}f(e^{i\frac{2\pi}{3}n}x,\alpha,\{s_{k+2n}\}),
\end{eqnarray}
where the bar means the complex conjugation.

\subsection{Asymptotic solution for $s_2=0$}

Let us consider the RH problem above where $s_2=0$ assuming that
$|x|\to\infty$ in the sector $\arg x\in[0,\frac{\pi}{3}]$. Equations
(\ref{p24}) imply that $s_5=0$ as well, and the function $Y(\lambda)$ defined
in (\ref{Y_expansion}) has no jump across two rays
$\arg\lambda=\frac{\pi}{2},\frac{3\pi}{2}$, $|\lambda|>r$. The remaining
parameters $s_1$ and $s_3$ satisfy the constraint
\begin{equation}\label{s13_relation}
s_1+s_3=-2\sin\pi\alpha,
\end{equation}
and the connection matrix $E$ due to (\ref{E_conditions}) has the form
\begin{equation}\label{E0}
E=\pmatrix{p&\cr&q\cr}
\pmatrix{1&ie^{-i\pi\alpha}\cr1&-ie^{i\pi\alpha}\cr},\quad
pq=-\frac{1}{2i\cos\pi\alpha}.
\end{equation}

Our first step in the RH problem analysis is elementary and consists of the
formulation of the equivalent RH problem for the piece-wise holomorphic
function $Y(\lambda)=\Psi(\lambda)e^{\theta(\lambda)\sigma_3}$:
\begin{eqnarray}\label{Y_RH}
i)&&Y(\lambda)\to I\quad\hbox{as}\quad
\lambda\to\infty,\quad
\|Y(\lambda)\lambda^{-\alpha\sigma_3}\|\leq const\quad\hbox{as}\quad
\lambda\to0;\nonumber
\\
ii)&&Y_+(\lambda)=Y_-(\lambda)G(\lambda),\quad
G(\lambda)=e^{-\theta\sigma_3}S(\lambda)e^{\theta\sigma_3},\quad
\lambda\in\gamma.
\end{eqnarray}

If $S(\lambda)=I+s\sigma_{\pm}$ then
$G(\lambda)=I+se^{\mp2\theta}\sigma_{\pm}$. On the second step, we transform
the jump contour $\gamma$ to the contour of the steepest descent for the
matrix $G(\lambda)-I$.

Let us denote by $\gamma_+$ the level line $\Im\theta(\lambda)=const$ passing
through the stationary phase point $\lambda_+=\frac{i}{2}x^{1/2}$ and
asymptotic to the rays $\arg\lambda=\frac{\pi}{6},\frac{5\pi}{6}$. This is
the steepest descent line for $e^{2\theta}$. Similarly, let us denote
by $\gamma_-$ the level line passing through the stationary phase point
$\lambda_-=-\frac{i}{2}x^{1/2}$ and asymptotic to the rays
$\arg\lambda=-\frac{\pi}{6},-\frac{5\pi}{6}$. This is the steepest descent
line for $e^{-2\theta}$.

For $\arg x\in(0,\frac{\pi}{3}]$, the level lines $\ell_+$ and $\ell_-$
emanating from the origin and asymptotic to the respective rays
$\arg\lambda=\frac{\pi}{6}$ and $\arg\lambda=\frac{5\pi}{6}$ are the
steepest descent lines for $e^{2\theta}$ and $e^{-2\theta}$, correspondingly.

For $\arg x=0$, the level lines $\ell_+$ and $\ell_-$ are the segments
connecting the origin $\lambda=0$ with the respective stationary phase points
$\lambda_+$ and $\lambda_-$. For $x>4r^2$, these lines contain the radiuses
$\rho_+$ and $\rho_-$.

It is convenient to consider the following equivalent RH problems for
$\Psi(\lambda)$:

\noindent
for $\arg x\in(0,\frac{\pi}{3}]$, the jump contour is the union of the level
lines $\gamma_+$, $\gamma_-$, $\ell_+$, $\ell_-$ oriented from the left to
the right, and of the circle $C=C_r\cup C_l$ divided by $\ell_{\pm}$ into the
right $C_r$ and left $C_l$ arcs both clockwise oriented. The jump matrices
are as follows:
\begin{eqnarray}\label{+jumps}
&&\hskip-20pt
\lambda\in\gamma_+\colon\
S(\lambda)=S_3^{-1}=\pmatrix{1&0\cr-s_3&1\cr},\quad
\lambda\in\gamma_-\colon\
S(\lambda)=S_6=\pmatrix{1&-s_3\cr0&1\cr},\nonumber
\\
&&\hskip-20pt
\lambda\in\ell_+,\
|\lambda|>r\colon\
S(\lambda)=S_1S_3=\pmatrix{1&0\cr-2\sin\pi\alpha\,&1\cr}=S_+,\nonumber
\\
&&\hskip-20pt
\lambda\in\ell_-,\
|\lambda|>r\colon\
S(\lambda)=S_6^{-1}S_4^{-1}=\pmatrix{1&-2\sin\pi\alpha\cr0&1\cr}=S_-,\nonumber
\\
&&\hskip-20pt
\lambda\in\ell_+,\
|\lambda|<r\colon\
S(\lambda)=\sigma_2M^{-1}\sigma_2,\quad
\lambda\in\ell_-,\
|\lambda|<r\colon\
S(\lambda)=M,\nonumber
\\
&&\hskip-20pt
\lambda\in C_r\colon\
S(\lambda)=E,\quad
\lambda\in C_l\colon\
S(\lambda)=\sigma_2E\sigma_2.
\end{eqnarray}

The jump contour for the RH problem (\ref{+jumps}) is decomposed into the
disjoint sum of the lines $\gamma_+$, $\gamma_-$ and
$\gamma_0=\ell_+\cup\ell_-\cup C$. For the boundary value $\arg x=0$, the
lines $\ell_+$ and $\ell_-$ emanating from the origin pass through the
stationary phase points and partially merge with the lines $\gamma_+$ and
$\gamma_-$. Thus the corresponding RH problem graph can not be decomposed
into a disjoint union of the steepest descent contours except for
$\alpha\in{\Bbb Z}$ when the jump across $\gamma_0$ can be eliminated. The
particular case $\alpha=0$ was studied in \cite{FN, its_nov}. The general
case is described in the following way:

\noindent
for $\arg x=0$, the jump contour is the union of the level lines $\gamma_+$,
$\gamma_-$ oriented from the left to the right, of the level lines $\ell_+$
and $\ell_-$ oriented from $\lambda_-$ to $\lambda_+$, and of the
clockwise oriented circle $C=C_r\cup C_l$. The jump matrices are as
follows:
\begin{eqnarray}\label{0jumps}
&&\hskip-9pt
\lambda\in\gamma_+\colon\
S(\lambda)=\cases{S_3^{-1},\
\Re\lambda<0,\cr
S_1,\
\Re\lambda>0,\cr}\quad
\lambda\in\gamma_-\colon\
S(\lambda)=
\cases{S_4^{-1},\
\Re\lambda<0,\cr
S_6,\
\Re\lambda>0,\cr}\nonumber
\\
&&\hskip-9pt
\lambda\in\ell_+,\
|\lambda|>r\colon\
S(\lambda)=S_1S_3=\pmatrix{1&0\cr-2\sin\pi\alpha&1\cr},\nonumber
\\
&&\hskip-9pt
\lambda\in\ell_-,\
|\lambda|>r\colon\
S(\lambda)=S_6^{-1}S_4^{-1}=\pmatrix{1&-2\sin\pi\alpha\cr0&1\cr},\nonumber
\\
&&\hskip-9pt
\lambda\in\ell_+,\
|\lambda|<r\colon\
S(\lambda)=\sigma_2M^{-1}\sigma_2,\quad
\lambda\in\ell_-,\
|\lambda|<r\colon\
S(\lambda)=M,\nonumber
\\
&&\hskip-9pt
\lambda\in C_r\colon\
S(\lambda)=E,\quad
\lambda\in C_l\colon\
S(\lambda)=\sigma_2E\sigma_2.
\end{eqnarray}

The problem (\ref{0jumps}) is the limiting case $\arg x\to0$ of
(\ref{+jumps}).

As $\arg x\in[0,\frac{\pi}{3}]$, introduce the reduced RH problem
($s_2=s_3=0$) for the piece-wise holomorphic function $\Phi(\lambda)$ on
$\gamma_0=\ell_+\cup\ell_-\cup C$ oriented as above:
\begin{eqnarray}\label{Phi_RH}
i)&&\Phi(\lambda)e^{\theta\sigma_3}\to I,\quad
\lambda\to\infty,\qquad
\|\Phi(\lambda)\lambda^{-\alpha\sigma_3}\|\leq const,\quad
\lambda\to0,\nonumber
\\
ii)&&\Phi_+(\lambda)=\Phi_-(\lambda)S(\lambda),\quad
\lambda\in\gamma_0,
\end{eqnarray}
where the jump matrix $S(\lambda)$ is described in (\ref{+jumps}).

\begin{theorem}\label{Theorem1}
If $\alpha-\frac{1}{2}\notin{\Bbb Z}$ and $\arg x\in[0,\frac{\pi}{3}]$ while
$|x|$ is large enough, then there exists a unique solution of the RH problem
(\ref{Phi_RH}).
\end{theorem}
\proof
Since $\det S(\lambda)\equiv1$, we have $\det\Phi_+=\det\Phi_-$, and hence
$\det\Phi(\lambda)$ is an entire function. Furthermore, because of the
normalization of $\Phi(\lambda)$ at infinity, $\det\Phi(\lambda)\equiv$. Let
$\tilde\Phi$ and $\Phi$ be two solutions of (\ref{Phi_RH}). Then
$\chi(\lambda)=\tilde\Phi(\lambda)\Phi^{-1}(\lambda)$ is a rational function
of $\lambda$ with the only possible pole at $\lambda=0$. However, due to the
boundedness of $\Phi(\lambda)\lambda^{-\alpha\sigma_3}$ and
$\tilde\Phi(\lambda)\lambda^{-\alpha\sigma_3}$ as $\lambda\to0$, we have
$\chi(\lambda)$ bounded at $\lambda=0$ and, using the Liouville theorem,
$\chi(\lambda)\equiv const$. Thus, because of the normalization of $\Phi$ and
$\tilde\Phi$ at infinity, $\Phi(\lambda)\equiv\tilde\Phi(\lambda)$, and the
solution is unique.

To prove the existence of the solution $\Phi(\lambda)$, introduce an
auxiliary function
\begin{equation}\label{hat_Phi_0r}
\hat\Phi^0(z)=
B(z)\pmatrix{v_1(z)&v_2(z)\cr v_1'(z)&v_2'(z)\cr},\quad
B(z)=\frac{1}{2}e^{-i\frac{\pi}{4}\sigma_3}
\pmatrix{1&1\cr-1&1\cr}
\pmatrix{1&0\cr-\frac{\alpha}{z}&1\cr},
\end{equation}
where the prime means differentiation w.r.t.\ $z$ and, for
$\alpha-\frac{1}{2}\notin{\Bbb Z}$,
\begin{eqnarray}\label{y12}
&&\hskip-20pt
v_1(z)=\sum_{k=0}^{\infty}
\frac{\Gamma(\alpha+\frac{1}{2})z^{\alpha+2k}}
{4^kk!\Gamma(\alpha+\frac{1}{2}+k)}=
2^{\alpha-\frac{1}{2}}\Gamma(\alpha+\frac{1}{2})
e^{i\frac{\pi}{2}(\alpha-\frac{1}{2})}z^{\frac{1}{2}}
J_{\alpha-\frac{1}{2}}(-iz),\nonumber
\\
&&\hskip-20pt
v_2(z)=\sum_{k=0}^{\infty}
\frac{\Gamma(\frac{3}{2}-\alpha)z^{1-\alpha+2k}}
{4^kk!\Gamma(\frac{3}{2}-\alpha+k)}=
2^{\frac{1}{2}-\alpha}\Gamma(\frac{3}{2}-\alpha)
e^{i\frac{\pi}{2}(\frac{1}{2}-\alpha)}
z^{\frac{1}{2}}J_{\frac{1}{2}-\alpha}(-iz).
\end{eqnarray}
Here $J_{\nu}(z)$ denote the classical Bessel function \cite{BE}. It is worth
to note that the function $\hat\Phi^0(z)$ satisfies the linear differential
equation
\begin{equation}\label{Phi0_eq}
\hat\Phi^0_z=\bigl(\sigma_3-\frac{\alpha}{z}\sigma_2\bigr)\hat\Phi^0.
\end{equation}
As it is easily seen, if $|z|<const$, then
\begin{equation}\label{hat_Phi_0}
\|\hat\Phi^0(z)z^{-\alpha\sigma_3}\|\leq const.
\end{equation}
Using the properties of the Bessel functions, we find that the products
\begin{equation}\label{hat_Phi_k}
\hat\Phi_1(z)=\hat\Phi^0(z)\hat E,\quad
\hat\Phi_2(z)=\hat\Phi_1(z)\hat S_1,\quad
\hat\Phi_3(z)=\hat\Phi_2(z)\hat S_2,
\end{equation}
where
$$
\hat E=\frac{\sqrt\pi}{2\cos\pi\alpha}
\pmatrix{\frac{2^{1-\alpha}}{\Gamma(\frac{1}{2}+\alpha)}&\cr
&\frac{2^{\alpha}}{\Gamma(\frac{3}{2}-\alpha)}\cr}
e^{i\frac{\pi}{4}\sigma_3}
\pmatrix{e^{-i\pi\alpha}&i\cr ie^{i\pi\alpha}&1\cr},$$
$$
\hat S_1=\pmatrix{1&2\sin\pi\alpha\cr0&1\cr},\quad
\hat S_2=\pmatrix{1&0\cr-2\sin\pi\alpha&1\cr},$$
satisfy the asymptotic relation
\begin{equation}\label{hat_Phi_k_as}
\hat\Phi_k(z)=
(I-\frac{i\alpha}{2z}\sigma_1+{\cal O}(\frac{1}{z^2}))e^{z\sigma_3},\quad
z\to\infty,\quad
\arg z\in\bigl(\pi(k-\frac{3}{2}),\pi(k+\frac{1}{2})\bigr).
\end{equation}
We observe also the symmetries,
\begin{equation}\label{hat_symm}
\sigma_2\hat\Phi^0(e^{i\pi}z)\sigma_2=
\hat\Phi^0(z)M,\quad
M=ie^{i\pi\alpha\sigma_3}\sigma_1,\quad
\sigma_2\hat\Phi_{k+1}(e^{i\pi}z)\sigma_2=
\hat\Phi_k(z),
\end{equation}
and the relation
\begin{equation}\label{cyclic}
\hat E\hat S_1=DE,
\end{equation}
$$
D=\frac{\sqrt\pi\,e^{i\frac{\pi}{4}}}{2\cos\pi\alpha}
\pmatrix{\frac{2^{1-\alpha}e^{-i\pi\alpha}}{p\Gamma(\frac{1}{2}+\alpha)}&\cr
&\frac{2^{\alpha}e^{i\pi\alpha}}{q\Gamma(\frac{3}{2}-\alpha)}\cr},\quad
E=\pmatrix{p&\cr&q\cr}
\pmatrix{1&ie^{-i\pi\alpha}\cr1&-ie^{i\pi\alpha}\cr}.$$

Introduce the graph $\hat\gamma_0={\Bbb R}\cup\hat C$ consisting of the real
axis oriented from the left to the right and of the clockwise oriented
circle $\hat C=\{z\in{\Bbb C}\colon\ |z|=\hat r\}$ divided by the real axis
in the lower $\hat C_d$ and upper $\hat C_u$ arcs. This graph divides the
complex $z$-plane into four regions: $\hat\Omega_2$ which is the exterior of
the circle in the lower half of the complex $z$-plane, $\hat\Omega_3$ which
is the exterior of the circle in the upper half of the complex $z$-plane,
$\hat\Omega_d$ which is the lower half of the interior of the circle and
$\hat\Omega_u$ which is the upper half of the interior of the circle. Define
a piece-wise holomorphic function $\hat\Phi(z)$,
\begin{equation}\label{hat_Phi_collect}
\hat\Phi(z)=\cases{
\hat\Phi_2(z),\quad
z\in\hat\Omega_2,\cr
\hat\Phi_3(z),\quad
z\in\hat\Omega_3,\cr
\hat\Phi^0(z)D,\quad
z\in\hat\Omega_d,\cr
\sigma_2\hat\Phi^0(e^{-i\pi}z)D\sigma_2,\quad
z\in\hat\Omega_u.\cr}
\end{equation}
By construction, this function solves the following RH problem:
\begin{eqnarray}\label{hat_Phi_as_RH}
i)&&
\hat\Phi(z)e^{-z\sigma_3}\to I,\quad
z\to\infty,\qquad
\|\hat\Phi(z)z^{-\alpha\sigma_3}\|\leq const,\quad
z\to0;
\\
ii)&&\hbox{on the contour $\hat\gamma_0$, the jump condition
$\hat\Phi_+(z)=\hat\Phi_-(z)\hat S(z)$ holds true:}\nonumber
\\\label{hat_jumps}
&&z>\hat r\colon\quad
\hat S(z)=\hat S_2=\pmatrix{1&0\cr-2\sin\pi\alpha&1\cr}=S_+,\nonumber
\\
&&z<-\hat r\colon\quad
\hat S(z)=\hat S_1^{-1}=\pmatrix{1&-2\sin\pi\alpha\cr0&1\cr}=S_-,\nonumber
\\
&&-\hat r<z<0\colon\quad
\hat S(z)=M,\qquad
0<z<\hat r\colon\quad
\hat S(z)=\sigma_2M^{-1}\sigma_2,\nonumber
\\
&&z\in\hat C_d\colon\quad
\hat S(z)=E,\qquad
z\in\hat C_u\colon\quad
\hat S(z)=\sigma_2E\sigma_2.
\end{eqnarray}

Therefore the function $\hat\Phi(z)$ has precisely the jump properties of the
function $\Phi(\lambda)$. To find $\Phi(\lambda)$ with the proper asymptotic
condition at infinity, let us replace the jump contour $\hat\gamma_0$ by the
contour obtained from $\gamma_0$ using the mapping
\begin{equation}\label{z_lambda}
z(\lambda)=-\theta(\lambda)=-ix\lambda-i\frac{4}{3}\lambda^3.
\end{equation}
For $|\lambda|\leq R<\frac{1}{2}|x|^{1/2}$, the mapping
(\ref{z_lambda}) gives us the holomorphic change of the independent variable.
Introduce the piece-wise holomorphic function $\tilde\Phi(\lambda)$,
\begin{equation}\label{tilde_Phi}
\tilde\Phi(\lambda)=\cases{
\hat\Phi(z(\lambda)),\quad
|\lambda|<R,\cr
e^{-\theta(\lambda)\sigma_3},\quad
|\lambda|>R.\cr}
\end{equation}
We look for the solution of the RH problem (\ref{Phi_RH}) in the form of the
product
\begin{equation}\label{chi_def}
\Phi(\lambda)=\chi(\lambda)\tilde\Phi(\lambda).
\end{equation}
With this purpose, consider the RH problem for the correction function
$\chi(\lambda)$ on the union $\ell$ of the clockwise oriented circle
${\cal L}$ of the radius $R$ divided by the level lines $\ell_+$ and $\ell_-$
into the left ${\cal L}_l$ and right ${\cal L}_r$ arcs and of the outer parts
of the lines $\ell_+$, $\ell_-$:
\begin{eqnarray}\label{chi_RH}
i)&&\chi(\lambda)\to I,\quad
\lambda\to\infty;\nonumber
\\
ii)&&\chi^+(\lambda)=\chi^-(\lambda)H(\lambda),\quad
\lambda\in\ell,
\\
&&\lambda\in\ell_+,\
|\lambda|>R\colon\quad
H(\lambda)=e^{-\theta\sigma_3}S_+e^{\theta\sigma_3},
\nonumber
\\
&&\lambda\in\ell_-,\
|\lambda|>R\colon\quad
H(\lambda)=e^{-\theta\sigma_3}S_-e^{\theta\sigma_3},
\nonumber
\\
&&|\lambda|=R,\
\lambda\in{\cal L}_r\colon\quad
H(\lambda)=\hat\Phi_2(z(\lambda))e^{\theta\sigma_3},
\nonumber
\\
&&|\lambda|=R,\
\lambda\in{\cal L}_l\colon\quad
H(\lambda)=\hat\Phi_3(z(\lambda))e^{\theta\sigma_3}.
\end{eqnarray}
Taking into account the equations (\ref{hat_jumps}), $S_+=\hat S_2$,
$S_-=\hat S_1^{-1}$, it is easy to see the continuity of the RH problem at
the node points $\ell_+\cap{\cal L}$ and $\ell_-\cap{\cal L}$. Furthermore,
the jump matrix $H(\lambda)$ satisfies the estimates
\begin{equation}\label{H_jump_estims}
\|H(\lambda)-I\|,\|\frac{\partial H}{\partial\lambda}\|\leq
\cases{c_1(|x|+|\lambda|^2)e^{-c_2|x|\,|\lambda|},\quad
\lambda\in\ell_{\pm},\quad
|\lambda|\geq R,\cr
c_1R^{-3},\quad
|\lambda|=R,\cr}
\end{equation}
where the concrete value of the constants $c_j$, $j=1,2$, is not important
for us. Because we may take $R=c|x|^{1/2}$ with some positive constant $c$,
the solvability of the RH problem (\ref{chi_RH}) and therefore of
(\ref{Phi_RH}) is straightforward. Indeed, consider the equivalent system of
the non-homogeneous singular integral equations for the limiting value
$\chi^-(\lambda)$, i.e.\
\begin{equation}\label{chi_sing}
\chi^-(\lambda)=I+\frac{1}{2\pi i}\int_{\ell}
\frac{\chi^-(\zeta)(H(\zeta)-I)}{\zeta-\lambda_-}\,d\zeta,
\end{equation}
or, in the symbolic form, $\chi^-=I+K\chi^-$. Here $\lambda_-$ means the
right limit of $\lambda$ on $\ell$, and $K$ is the composition of the
operator of the right multiplication in $H-I$ and of the Cauchy operator
$C_-$. An equivalent singular integral equation for $\psi^-:=\chi^--I$
differs from (\ref{chi_sing}) in the nonhomogeneous term only,
\begin{equation}\label{psi_sing}
\psi^-=KI+K\psi^-.
\end{equation}
Consider the integral equation (\ref{psi_sing}) in the space $H^1(\ell)$ of
the functions $\psi^-$ such that $\psi^-$ and its distributional derivative
both belong to $L^2(\ell)$. Since $H-I$ is small in $H^1(\ell)$ for large
enough $|x|$, and $C_-$ is bounded in $H^1(\ell)$ \cite{zhou}, then
$\|K\|_{H^1(\ell)}\leq c|x|^{-3/2}$ with some constant $c$, while $I-K$ is
invertible in $H^1(\ell)$. Because $KI\in H^1(\ell)$, equation
(\ref{psi_sing}) for $\psi^-$ is solvable in $H^1(\ell)$, and the solution
$\chi(\lambda)$ of the RH problem (\ref{chi_RH}) is determined by
$\psi^-(\lambda)$ using the equation $\chi=I+KI+K\psi^-$.
\endproof

\smallskip
In accord with the said above, the function $\chi^-$ is given by the
converging iterative series, $\chi^-=\sum_{n=0}^{\infty}K^nI$. Thus, for
large enough $|x|$ and $\arg x\in[0,\frac{\pi}{3}]$, the Painlev\'e function
$y_1(x)$ corresponding to $s_2=s_3=0$, $s_1=-2\sin\pi\alpha$, is given by
(\ref{y_from_Y}),
\begin{equation}\label{y0_def}
y_1(x)=2\bigl(\lim_{\lambda\to\infty}\chi(\lambda)\lambda\bigr)_{12}=
-\frac{1}{\pi i}
\int_{\ell}\bigl(\chi^-(\zeta)(H(\zeta)-I)\bigr)_{12}\,d\zeta.
\end{equation}
Since $\chi^-(\zeta)=I+{\cal O}(\zeta^{-1})$ and $|\zeta|\geq R=c|x|^{1/2}$,
we find
$$
y_1(x)=-\frac{1}{\pi i}
\int_{\ell}(H(\lambda)-I)_{12}\,d\lambda\cdot(I+{\cal O}(x^{-1/2})).$$
Using here the expressions for the jump matrix $H(\lambda)$ from
(\ref{chi_RH}), the asymptotics (\ref{hat_Phi_k_as}), and the definition
(\ref{z_lambda}), we find the leading asymptotic term of $y_1(x)$,
\begin{equation}\label{y0_as}
y_1(x)=\frac{\alpha}{x}+{\cal O}(x^{-3/2}),\quad
\arg x\in[0,\frac{\pi}{3}].
\end{equation}

Let us go to the case of the nontrivial $s_3$ described by the RH problem
(\ref{+jumps}). We look for the solution $\Psi(\lambda)$ in the form of the
product
\begin{equation}\label{Psi_chi}
\Psi(\lambda)=X(\lambda)\Phi(\lambda),
\end{equation}
where $\Phi(\lambda)$ is the solution of the reduced RH problem
(\ref{Phi_RH}). Thus we arrive at the RH problem for the correction function
$X(\lambda)$,
\begin{eqnarray}\label{Psi_X_RH}
i)&&X(\lambda)\to I,\quad
\lambda\to\infty,\nonumber
\\
ii)&&X_+(\lambda)=X_-(\lambda){\cal H}(\lambda),\quad
\lambda\in\gamma_+\cup\gamma_-,
\\
&&\lambda\in\gamma_+\colon\quad
{\cal H}(\lambda)=\Phi_-(\lambda)S_3^{-1}\Phi_-^{-1}(\lambda),\nonumber
\\
&&\lambda\in\gamma_-\colon\quad
{\cal H}(\lambda)=\Phi_-(\lambda)S_6\Phi_-^{-1}(\lambda),\nonumber
\end{eqnarray}
where $\Phi_-(\lambda)$ denotes the right limit of $\Phi(\lambda)$ on
$\gamma_+\cup\gamma_-$.

The estimate for the jump matrix on $\gamma_{\pm}$,
\begin{equation}\label{X_H_estim}
\|{\cal H}(\lambda)-I\|\leq ce^{-\frac{2}{3}|x|^{3/2}\cos(\frac{3}{2}\arg x)}
e^{-4|x|^{1/2}|\lambda-\lambda_0|^2},
\end{equation}
where $c$ is some constant and $\lambda_0=\pm\frac{i}{2}x^{1/2}$ is the
stationary phase point, yields the estimate for the norm of the singular
integral operator ${\cal K}$ in the equivalent system of singular integral
equations, $X_-=I+{\cal K}X_-$,
\begin{equation}\label{K_estim}
\|{\cal K}\|_{L^2(\gamma_+\cup\gamma_-)}\leq
ce^{-\frac{2}{3}|x|^{3/2}\cos(\frac{3}{2}\arg x)}.
\end{equation}
If $|x|$ is large enough and $\arg x\in[0,\frac{\pi}{3}-\epsilon]$,
$\epsilon>0$, then the operator ${\cal K}$ is contracting and the system
$X_-=I+{\cal K}X_-$ is solvable by iterations in
$L^2(\gamma_+\cup\gamma_-)$, i.e.\
$X_-=\sum_{n=0}^{\infty}{\cal K}^nX_-=I+{\cal K}I+
{\cal O}(e^{-\frac{4}{3}x^{3/2}})$. However, to incorporate the oscillating
direction $\arg x=\frac{\pi}{3}$ in the general scheme, we use some more
refined procedure.

\begin{theorem}\label{Theorem2}
If $s_2=0$ and $|x|\to\infty$, $\arg x\in[0,\frac{\pi}{3}]$, then the
asymptotics of the second Painlev\'e transcendent is given by
\begin{equation}\label{y+_as}
y=y_1(x,\alpha)-\frac{is_3}{2\sqrt\pi}\,x^{-1/4}e^{-\frac{2}{3}x^{3/2}}
(1+{\cal O}(x^{-1/2}))+{\cal O}(s_3^2e^{-\frac{4}{3}x^{3/2}}),
\end{equation}
where $y_1(x,\alpha)\sim\frac{\alpha}{x}$ is the solution of the Painlev\'e
equation for $s_2=s_3=0$, $s_1=-2\sin\pi\alpha$.
\end{theorem}

\proof
Using the asymptotics of $\Phi(\lambda)$ at infinity (\ref{Phi_RH}), we find
the asymptotics of the jump matrix ${\cal H}(\lambda)$,
\begin{equation}\label{H_as}
{\cal H}(\lambda)=\cases{
I-s_3e^{2\theta}\sigma_-+{\cal O}(\psi^-e^{2\theta}),\quad
\lambda\in\gamma_+,\cr
I-s_3e^{-2\theta}\sigma_++{\cal O}(\psi^-e^{-2\theta}),\quad
\lambda\in\gamma_-,\cr}
\end{equation}
where $\psi^-=(I-K)^{-1}KI$ introduced above is holomorphic in a neighborhood
of $\gamma_{\pm}$ if $\arg x\in(0,\frac{\pi}{3}]$ and
$\psi^-\in H^1(\gamma_+\cup\gamma_-)$ if $\arg x=0$ and satisfies the
estimate
$$
\|\psi^-(\lambda)\|\leq c(|x|^{3/2}+|\lambda|)^{-1},\quad
\lambda\in\gamma_{\pm},$$
with some constant $c$ which value is not important for us.

Consider the triangular matrices $P(\lambda)$, $Q(\lambda)$,
\begin{eqnarray}\label{PQ_def}
&&P(\lambda)=I-\frac{s_3}{2\pi i}
\int_{\gamma_+}e^{2\theta(\zeta)}\frac{d\zeta}{\zeta-\lambda}\,\sigma_-,
\nonumber
\\
&&Q(\lambda)=I-\frac{s_3}{2\pi i}
\int_{\gamma_-}e^{-2\theta(\zeta)}\frac{d\zeta}{\zeta-\lambda}\,\sigma_+,
\end{eqnarray}
which solve the triangular RH problems:

i) $P(\lambda)\to I$ as $\lambda\to\infty$, and
$P_+(\lambda)=P_-(\lambda)(I-s_3e^{2\theta}\sigma_-)$ as $\lambda\in\gamma_+$,

ii) $Q(\lambda)\to I$ as $\lambda\to\infty$, and
$Q_+(\lambda)=Q_-(\lambda)(I-s_3e^{-2\theta}\sigma_+)$ as
$\lambda\in\gamma_-$.

We will look for solution $X(\lambda)$ of the RH problem (\ref{Psi_X_RH}) in
the form of the product
\begin{equation}\label{X_PQ_Z}
X(\lambda)=Z(\lambda)Q(\lambda)P(\lambda).
\end{equation}
The correction function $Z(\lambda)$ satisfies the RH problem
\begin{eqnarray}\label{Z_RH}
i)&&Z(\lambda)\to I,\quad
\lambda\to\infty,\nonumber
\\
ii)&&Z_+(\lambda)=Z_-(\lambda)V(\lambda),\quad
\lambda\in\gamma_+\cup\gamma_-,
\end{eqnarray}
where
\begin{eqnarray}\label{V}
&&V(\lambda)=Q(\lambda)P_-(\lambda){\cal H}(\lambda)(I+s_3e^{2\theta}\sigma_-)
P_-^{-1}(\lambda)Q^{-1}(\lambda),\quad
\lambda\in\gamma_+,\nonumber
\\
&&V(\lambda)=Q_-(\lambda)P(\lambda){\cal H}(\lambda)P^{-1}(\lambda)
(I+s_3e^{-2\theta}\sigma_+)Q_-^{-1}(\lambda),\quad
\lambda\in\gamma_-.
\end{eqnarray}
Using the first of the relations in (\ref{H_as}), we find the estimate for
the jump matrix $V(\lambda)$ on $\gamma_+$,
\begin{equation}\label{V_as+}
V(\lambda)=I+{\cal O}(\psi^-e^{2\theta})+
{\cal O}(\phi(\lambda)e^{2\theta}),\quad
\lambda\in\gamma_+,
\end{equation}
where $\phi(\lambda)=
\int_{\gamma_-}e^{-2\theta(\zeta)}\frac{d\zeta}{\zeta-\lambda}$ is square
integrable and holomorphic in a neighborhood of $\gamma_+$. This function
satisfies the estimate
$$
|\phi(\lambda)|\leq
c(|x|^{1/2}+|\lambda|)^{-1}
e^{-\frac{2}{3}|x|^{3/2}\cos(\frac{3}{2}\arg x)},\quad
\lambda\in\gamma_+,$$
with some constant $c$ which value is not important for us. For the product
$P{\cal H}P^{-1}$ on $\gamma_-$, we compute the expression
$$
P(\lambda){\cal H}(\lambda)P^{-1}(\lambda)=I-s_3e^{-2\theta}\sigma_++
{\cal O}(\varphi(\lambda)e^{-2\theta})+{\cal O}(\psi^-e^{-2\theta}),\quad
\lambda\in\gamma_-,$$
where $\varphi(\lambda)=
\int_{\gamma_+}e^{2\theta(\zeta)}\frac{d\zeta}{\zeta-\lambda}$ is square
integrable and holomorphic in a neighborhood of $\gamma_-$. This function
satisfies the estimate
$$
|\varphi(\lambda)|\leq
c(|x|^{1/2}+|\lambda|)^{-1}
e^{-\frac{2}{3}|x|^{3/2}\cos(\frac{3}{2}\arg x)},\quad
\lambda\in\gamma_-,$$
with some constant $c$. Thus,
\begin{equation}\label{V_as-}
V(\lambda)=I+{\cal O}(\psi^-e^{-2\theta})+{\cal O}(\varphi e^{-2\theta}),\quad
\lambda\in\gamma_-.
\end{equation}

Our next steps are similar to presented in the proof of
Theorem~\ref{Theorem1}. Consider the system of the singular integral
equations for $Z_-(\lambda)$ equivalent to the RH problem (\ref{Z_RH}),
$Z_-=I+{\Eu K}Z_-$. Here the singular integral operator ${\Eu K}$ is the
superposition of the multiplication operator in $V-I$ and of the Cauchy
operator $C_-$. Because the Cauchy operator is bounded in
$L^2(\gamma_+\cup\gamma_-)$, the singular integral operator ${\Eu K}$ for
large enough $|x|$, $\arg x\in[0,\frac{\pi}{3}]$, satisfies the estimate
\begin{equation}\label{Eu_K_estim}
\|{\Eu K}\|_{L^2(\gamma_+\cup\gamma_-)}\leq
c|x|^{-1/2}e^{-\frac{2}{3}|x|^{3/2}\cos(\frac{3}{2}\arg x)}.
\end{equation}
Thus equation $\zeta_-={\Eu K}I+{\Eu K}\zeta_-$ for the difference
$\zeta_-:=Z_--I$ is solvable by iterations in the space
$L^2(\gamma_+\cup\gamma_-)$. Solution of the RH problem (\ref{Z_RH}) is given
by the integral $Z=I+{\Eu K}I+{\Eu K}\zeta_-$. Thus, using (\ref{y_from_Y}),
(\ref{Psi_chi}), (\ref{X_PQ_Z}) and the asymptotics of $P$, $Q$ from
(\ref{PQ_def}), we arrive at (\ref{y+_as}).
\endproof

\subsection{The decreasing degenerate Painlev\'e functions}

Applying the second of the symmetries (\ref{P_symmetries}) to (\ref{y+_as}),
we obtain
\begin{theorem}\label{Theorem3}
If $s_2=0$ and $|x|\to\infty$, $\arg x\in[-\frac{\pi}{3},0]$, then the
asymptotics of the second Painlev\'e transcendent is given by
\begin{equation}\label{y-_as}
y=y_3(x,\alpha)+\frac{is_1}{2\sqrt\pi}\,x^{-1/4}e^{-\frac{2}{3}x^{3/2}}
(1+{\cal O}(x^{-1/2}))+{\cal O}(s_1^2e^{-\frac{4}{3}x^{3/2}}),
\end{equation}
where $y_3(x,\alpha)=\overline{y_1(\bar x,\bar\alpha)}\sim\frac{\alpha}{x}$
is the solution of the Painlev\'e equation for $s_2=s_1=0$,
$s_3=-2\sin\pi\alpha$.
\end{theorem}

The solutions $y_1(x,\alpha)$ and
$y_3(x,\alpha)=\overline{y_1(\bar x,\bar\alpha)}$ are meromorphic functions
of $x\in{\Bbb C}$ and thus can be continued beyond the sectors indicated in
Theorems~\ref{Theorem2} and~\ref{Theorem3}. To find the asymptotics of the
solution $y_3(x,\alpha)$ in the interior of the sector
$\arg x\in[0,\frac{\pi}{3}]$, we apply (\ref{y+_as}). Similarly, we find the
asymptotics of the solution $y_1(x,\alpha)$ in the interior of the sector
$\arg x\in[-\frac{\pi}{3},0]$ using (\ref{y-_as}). Both the expressions imply
that, if $|x|\to\infty$, $\arg x\in[-\frac{\pi}{3},\frac{\pi}{3}]$,
\begin{equation}\label{y0_hat_y0}
y_3(x,\alpha)-y_1(x,\alpha)=
i\frac{\sin\pi\alpha}{\sqrt\pi}x^{-1/4}e^{-\frac{2}{3}x^{3/2}}
(1+{\cal O}(x^{-1/2}))+{\cal O}(e^{-\frac{4}{3}x^{3/2}}).
\end{equation}

\smallskip
{\it Remark 1}. The same asymptotic relation can be obtained using the
analysis of the RH problem for the ratio
$\hat\Psi(\lambda)\Psi^{-1}(\lambda)$ where $\hat\Psi(\lambda)$ is the
solution of the RH problem for $s_2=s_1=0$ while $\Psi(\lambda)$ is the
solution for $s_2=s_3=0$.

\smallskip
Besides $y_1(x,\alpha)$ and $y_3(x,\alpha)$, let us introduce the solution
$y_2(x,\alpha)$ corresponding to the Stokes multipliers $s_1=s_3=0$,
$s_2=2\sin\pi\alpha$. Due to the second of the symmetries
(\ref{P_symmetries}), $y_2(x,\alpha)=\overline{y_2(\bar x,\bar\alpha)}$.

The last of the symmetries (\ref{P_symmetries}) yields the relations
\begin{eqnarray}\label{yk_rotate}
&&y_1(x,\alpha)=e^{i\frac{2\pi}{3}}y_2(e^{i\frac{2\pi}{3}}x,\alpha)=
e^{-i\frac{2\pi}{3}}y_3(e^{-i\frac{2\pi}{3}}x,\alpha),\nonumber
\\
&&y_2(x,\alpha)=e^{i\frac{2\pi}{3}}y_3(e^{i\frac{2\pi}{3}}x,\alpha)
=e^{-i\frac{2\pi}{3}}y_1(e^{-i\frac{2\pi}{3}}x,\alpha),\nonumber
\\
&&y_3(x,\alpha)=e^{i\frac{2\pi}{3}}y_1(e^{i\frac{2\pi}{3}}x,\alpha)
=e^{-i\frac{2\pi}{3}}y_2(e^{-i\frac{2\pi}{3}}x,\alpha),
\end{eqnarray}
which imply the decreasing asymptotics of the functions $y_k(x,\alpha)$ in
the following sectors:
\begin{eqnarray}\label{y_k_as_sectors}
&&y_1(x,\alpha)\sim\frac{\alpha}{x},\quad
|x|\to\infty,\quad
\arg x\in[0,\frac{2\pi}{3}],\nonumber
\\
&&y_2(x,\alpha)\sim\frac{\alpha}{x},\quad
|x|\to\infty,\quad
\arg x\in[\frac{2\pi}{3},\frac{4\pi}{3}],\nonumber
\\
&&y_3(x,\alpha)\sim\frac{\alpha}{x},\quad
|x|\to\infty,\quad
\arg x\in[-\frac{2\pi}{3},0].
\end{eqnarray}
The same symmetry (\ref{P_symmetries}) applied to the equation
(\ref{y0_hat_y0}) yields the differences
\begin{eqnarray}\label{yk_hat_yk}
&&\arg x\in[-\pi,-\frac{\pi}{3}]\colon\nonumber
\\
&&y_2(x,\alpha)-y_3(x,\alpha)=
-\frac{\sin\pi\alpha}{\sqrt\pi}x^{-1/4}e^{\frac{2}{3}x^{3/2}}
(1+{\cal O}(x^{-1/2}))+{\cal O}(e^{\frac{4}{3}x^{3/2}}),\nonumber
\\
&&\arg x\in[\frac{\pi}{3},\pi]\colon
\\
&&y_1(x,\alpha)-y_2(x,\alpha)=
\frac{\sin\pi\alpha}{\sqrt\pi}x^{-1/4}e^{\frac{2}{3}x^{3/2}}
(1+{\cal O}(x^{-1/2}))+{\cal O}(e^{\frac{4}{3}x^{3/2}}).\nonumber
\end{eqnarray}
Equations (\ref{y0_hat_y0}), (\ref{yk_hat_yk}) constitute the quasi-linear
Stokes phenomenon for the second Painlev\'e equation.

Finally, applying the last of the symmetries (\ref{P_symmetries}) to
(\ref{y+_as}), (\ref{y-_as}), and using (\ref{yk_rotate}) we find the
asymptotics of the degenerate Painlev\'e functions:

i) if $s_1=0$, then $s_3-s_2=-2\sin\pi\alpha$, and
\begin{eqnarray}\label{s1=0}
&&\hskip-20pt
\arg x\in[-\pi,-\frac{2\pi}{3}]\colon\
y=y_2(x,\alpha)-\frac{s_3}{2\sqrt\pi}\,x^{-1/4}e^{\frac{2}{3}x^{3/2}}
(1+{\cal O}(x^{-1/2})),\nonumber
\\
&&\hskip-20pt
\arg x\in[-\frac{2\pi}{3},-\frac{\pi}{3}]\colon\
y=y_3(x,\alpha)-\frac{s_2}{2\sqrt\pi}\,x^{-1/4}e^{\frac{2}{3}x^{3/2}}
(1+{\cal O}(x^{-1/2}));
\end{eqnarray}

ii) if $s_2=0$, then $s_1+s_3=-2\sin\pi\alpha$ and
\begin{eqnarray}\label{s2=0}
&&\hskip-20pt
\arg x\in[-\frac{\pi}{3},0]\colon\
y=y_3(x,\alpha)+\frac{is_1}{2\sqrt\pi}\,x^{-1/4}e^{-\frac{2}{3}x^{3/2}}
(1+{\cal O}(x^{-1/2})),\nonumber
\\
&&\hskip-20pt
\arg x\in[0,\frac{\pi}{3}]\colon\
y=y_1(x,\alpha)-\frac{is_3}{2\sqrt\pi}\,x^{-1/4}e^{-\frac{2}{3}x^{3/2}}
(1+{\cal O}(x^{-1/2}));
\end{eqnarray}

iii) if $s_3=0$, then $s_1-s_2=-2\sin\pi\alpha$ and
\begin{eqnarray}\label{s3=0}
&&\hskip-20pt
\arg x\in[\frac{\pi}{3},\frac{2\pi}{3}]\colon\
y=y_1(x,\alpha)-\frac{s_2}{2\sqrt\pi}\,x^{-1/4}e^{\frac{2}{3}x^{3/2}}
(1+{\cal O}(x^{-1/2})),\nonumber
\\
&&\hskip-20pt
\arg x\in[\frac{2\pi}{3},\pi]\colon\
y=y_2(x,\alpha)-\frac{s_1}{2\sqrt\pi}\,x^{-1/4}e^{\frac{2}{3}x^{3/2}}
(1+{\cal O}(x^{-1/2})).
\end{eqnarray}

\section{The power expansion of the degenerate solutions and the
coefficient asymptotics}

Using the steepest descent approach, cf.\ \cite{DZ2}, we can show the
existence of the asymptotic expansion of $y_k(x,\alpha)$, $k=1,2,3$, in the
negative degrees of $x^{1/2}$. Further elementary investigation of the
recursion relation for the coefficients of the series allows us to claim that
the asymptotic expansion for any of the decreasing degenerate solutions has
the following form:
\begin{equation}\label{y+formal}
y_0(x,\alpha)=\frac{\alpha}{x}\sum_{n=0}^{\infty}a_nx^{-3n}+
{\cal O}(x^{-\infty}),
\end{equation}
where coefficients $a_n$ are determined uniquely by the recurrence relation
\begin{equation}\label{an_recurrence}
a_0=1,\quad
a_{n+1}=(3n+1)(3n+2)a_n-2\alpha^2\sum_{k,l,m=0\atop k+l+m=n}^na_ka_la_m.
\end{equation}
The initial terms of the expansion are given by
\begin{eqnarray}\label{y0_expansion}
&&y_0(x,\alpha)=\frac{\alpha}{x}\Bigl\{1+
\frac{2(1-\alpha^2)}{x^3} +
\frac{4(10-13\alpha^2+3\alpha^4)}{x^6}+
\\[4pt]
&&+\frac{8}{x^9}\bigl(280-397\alpha^2+129\alpha^4-12\alpha^6\bigr)+\nonumber
\\[4pt]
&&+\frac{16}{x^{12}}\bigl(15400-22736\alpha^2+8427\alpha^4-
1146\alpha^6+55\alpha^8\bigr)+{\cal O}(x^{-15})\Bigr\}.\nonumber
\end{eqnarray}

For $\alpha=0$, the recurrence (\ref{an_recurrence}) is exactly solvable,
\begin{equation}\label{alpha0}
a_n=\frac{\Gamma(3n)}{3^{n-1}\Gamma(n)},\quad
\alpha=0.
\end{equation}
Our next goal is to determine the
asymptotics of the coefficients $a_n$ in (\ref{y+formal}) as $n\to\infty$ for
arbitrary $\alpha$. With this purpose, let us construct a sectorial analytic
function $\hat y(x)$,
\begin{equation}\label{hat_y}
\hat y(x)=\cases{y_3(x,\alpha),\quad
\arg x\in(-\frac{2\pi}{3},0),\cr
y_1(x,\alpha),\quad
\arg x\in(0,\frac{2\pi}{3}),\cr
y_2(x,\alpha),\quad
\arg x\in(\frac{2\pi}{3},\frac{4\pi}{3}).\cr}
\end{equation}
The function $\hat y(x)$ has a finite number of simple poles. Therefore
$\hat y(x)$ is bounded for $|x|\geq\rho$ and has the uniform expansion
(\ref{y+formal}) near infinity. Let $y^{(N)}(x)$ be a partial sum
\begin{equation}\label{yN}
y^{(N)}(x,\alpha)=\frac{\alpha}{x}\sum_{n=0}^{N-1}a_nx^{-3n},
\end{equation}
and $v^{(N)}(x)$ be a product
\begin{equation}\label{vn}
v^{(N)}(x)=x^{3N}(\hat y(x)-y^{(N)}(x,\alpha))=
\frac{\alpha}{x}\sum_{n=0}^{\infty}a_{n+N}x^{-3n}+{\cal O}(x^{-\infty}).
\end{equation}
Because $x^{3N}y^{(N)}(x,\alpha)$ is polynomial, the integral of $v^{(N)}(x)$
along the circle of the radius $|x|=\rho$ satisfies the estimate
\begin{equation}\label{vN_integral}
\Bigl|\oint_{|x|=\rho}v^{(N)}(x)\,dx\Bigr|\leq
\rho^{3N}\oint_{|x|=\rho}|\hat y(x)|\,dl\leq
2\pi\rho^{3N+1}\max_{|x|=\rho}|y(x)|=C\rho^{3N+1}.
\end{equation}
On the other hand, inflating the sectorial arcs of the circle $|x|=\rho$, we
find that
\begin{equation}\label{vN_inflation}
\oint_{|x|=\rho}v^{(N)}(x)\,dx=\oint_{|x|=R}v^{(N)}(x)\,dx+$$
$$
+\int_{\rho}^Rx^{3N}(y_1-y_3)\,dx+
\int_{e^{i\frac{2\pi}{3}}\rho}^{e^{i\frac{2\pi}{3}}R}x^{3N}(y_2-y_1)\,dx+
\int_{e^{-i\frac{2\pi}{3}}\rho}^{e^{-i\frac{2\pi}{3}}R}x^{3N}(y_3-y_2)\,dx.
\end{equation}
Because $v^{(N)}(x)=\frac{\alpha}{x}a_N+{\cal O}(x^{-4})$, the first of the
integrals in the r.h.s.\ of (\ref{vN_inflation}) is computed as follows:
\begin{equation}\label{aN_int}
\oint_{|x|=R}v^{(N)}(x)\,dx=2\pi i\alpha a_N+{\cal O}(R^{-3}).
\end{equation}
Last three integrals in (\ref{vN_inflation}) are computed using
(\ref{y0_hat_y0}), (\ref{yk_hat_yk}), (\ref{yk_rotate}):
$$
\int_{\rho}^Rx^{3N}(y_1-y_3)\,dx+
\int_{e^{i\frac{2\pi}{3}}\rho}^{e^{i\frac{2\pi}{3}}R}
x^{3N}(y_2-y_1)\,dx+
\int_{e^{-i\frac{2\pi}{3}}\rho}^{e^{-i\frac{2\pi}{3}}R}
x^{3N}(y_3-y_2)\,dx=$$
$$
=3\int_{\rho}^Rx^{3N}(y_1-y_3)\,dx=
-3i\frac{\sin\pi\alpha}{\sqrt\pi}
\int_{\rho}^Rx^{3N-\frac{1}{4}}e^{-\frac{2}{3}x^{3/2}}
(1+{\cal O}(x^{-1/2}))\,dx=$$
$$
=-3i\frac{\sin\pi\alpha}{\sqrt\pi}(\frac{3}{2})^{2N-\frac{1}{2}}\Bigl[
\Gamma(2N+\frac{1}{2})\bigl(1+{\cal O}(N^{-1/3})\bigr)+
{\cal O}(\rho^{3N})+{\cal O}(R^{3N}e^{-\frac{2}{3}R^{3/2}})\Bigr].$$
Thus, letting $R=\infty$, we find that the asymptotics as $N\to\infty$ of
the coefficient $a_N$ in (\ref{y+formal}) is given by
\begin{equation}\label{an_fin}
a_N=\frac{\sin\pi\alpha}{\alpha\pi^{3/2}}(3/2)^{2N+\frac{1}{2}}
\Gamma(2N+\frac{1}{2})\bigl(1+{\cal O}(N^{-1/3})\bigr)+{\cal O}(C^N),\quad
N\to\infty.
\end{equation}
For $\alpha=0$, this equation is consistent with equation (\ref{alpha0}).
For $\alpha\in{\Bbb Z}\backslash\{0\}$, the asymptotics (\ref{an_fin}) is
reduced to $a_N={\cal O}(C^N)$ which is consistent with the expansion of a
rational solution of P2 determined by the triviality condition
$s_1=s_2=s_3=0$ and $\alpha\in{\Bbb Z}$.

\bigskip
{\bf Acknowledgments.} The first co-author was supported in part by the
NSF, grant No.~DMS--9801608. The second co-author was supported in part by
the RFBR, grant No.~99--01--00687.

\ifx\undefined\bysame
\newcommand{\bysame}{\leavevmode\hbox to3em{\hrulefill}\,}
\fi

\end{document}